\documentclass[conference]{IEEEtran}
\IEEEoverridecommandlockouts
\usepackage{cite}
\usepackage{comment}
\usepackage{amsmath,amssymb,amsfonts}
\usepackage{algorithm}
\usepackage{algorithmic}

\usepackage{graphicx}
\usepackage{textcomp}
\usepackage{xcolor}
\usepackage{float}
\usepackage{multirow}
\usepackage{adjustbox}
\usepackage{cleveref}
\usepackage{subcaption}
\usepackage{flushend}
\usepackage[caption=false]{subfig}

\def\BibTeX{{\rm B\kern-.05em{\sc i\kern-.025em b}\kern-.08em
    T\kern-.1667em\lower.7ex\hbox{E}\kern-.125emX}}

\DeclareFontFamily{U}{stix2bb}{}
\DeclareFontShape{U}{stix2bb}{m}{n} {<-> stix2-mathbb}{}

\NewDocumentCommand{\indicator}{}{\text{\usefont{U}{stix2bb}{m}{n}1}}

\begin{document}

\title{Evaluating Query Efficiency and Accuracy of Transfer Learning-based Model Extraction Attack in Federated Learning}
\author{
    \IEEEauthorblockN{
    Sayyed Farid Ahamed\IEEEauthorrefmark{1}\textsuperscript{\textsection},
    Sandip Roy\IEEEauthorrefmark{1}\IEEEauthorrefmark{3}\textsuperscript{\textsection},
    Soumya Banerjee\IEEEauthorrefmark{1}\IEEEauthorrefmark{3},
    }
    \IEEEauthorblockN{
    Marc Vucovich\IEEEauthorrefmark{2}, 
    Kevin Choi\IEEEauthorrefmark{2}, 
    Abdul Rahman\IEEEauthorrefmark{2},
    Alison Hu\IEEEauthorrefmark{2},
    Edward Bowen\IEEEauthorrefmark{2},
    Sachin Shetty\IEEEauthorrefmark{1}
    }
    \IEEEauthorblockA{
    \IEEEauthorrefmark{1}Center for Secure \& Intelligent Critical Systems, Old Dominion University, Virginia, USA \\
    \IEEEauthorrefmark{3}School of Cybersecurity, Old Dominion University, Virginia, USA
    \\\{saham001,  sroy, s1banerj, sshetty\}@odu.edu}
    \IEEEauthorblockA{
    \IEEEauthorrefmark{2}Deloitte \& Touche LLP
    \\ mdvucovich@gmail.com, \{kevchoi, abdulrahman, aehu, edbowen\}@deloitte.com}
}

\maketitle
\begingroup\renewcommand\thefootnote{\textsection}
\footnotetext{The authors have equal contributions.}
\endgroup

\begin{abstract}
Federated Learning (FL) is a collaborative learning framework designed to protect client data, yet it remains highly vulnerable to Intellectual Property (IP) threats. Model extraction (ME) attack poses a significant risk to Machine-Learning-as-a-Service (MLaaS) platforms, enabling attackers to replicate confidential models by querying Black-Box (without internal insight) APIs. Despite FL’s privacy-preserving goals, its distributed nature makes it particularly susceptible to such attacks. 
This paper examines the vulnerability of the FL-based victim model to two types of model extraction attacks.
For various federated clients built under NVFlare platform, we implemented ME attack across two deep-learning architectures and three image datasets. We evaluate the proposed ME attack performance using various metrics, including accuracy, fidelity, and KL divergence. The experiments show that for various FL clients, the accuracy and fidelity of the extraction model are closely related to the size of the attack query set. Additionally, we explore a transfer learning-based approach where pre-trained models serve as the starting point for the extraction process. The results indicate that the accuracy and fidelity of the fine-tuned pre-trained extraction models are notably higher, particularly with smaller query sets, highlighting potential advantages for attackers.
\end{abstract}


\vspace{-10pt}
\section{Introduction}
Recently, Federated Learning (FL) has gained popularity as a privacy-focused machine learning (ML) approach that enables multiple clients to collaboratively create a consolidated model while safeguarding their training data \cite{mcmahan2016federated}. Unlike traditional centralized ML, FL eliminates the need for clients to transfer their raw data to a central server, thus protecting user privacy and security. The process typically involves training local models on client-specific data, sharing model updates among clients, and constructing a unified model accessible to all participants \cite{vucovich2022anomaly}. Since FL avoids data sharing, it effectively addresses privacy and security concerns commonly associated with centralized ML \cite{chen2021pois}, \cite{thakur2023novel}.

\begin{figure}[h!]
    \vspace{-9pt}
    \centering   
    \includegraphics[width=0.48\textwidth]{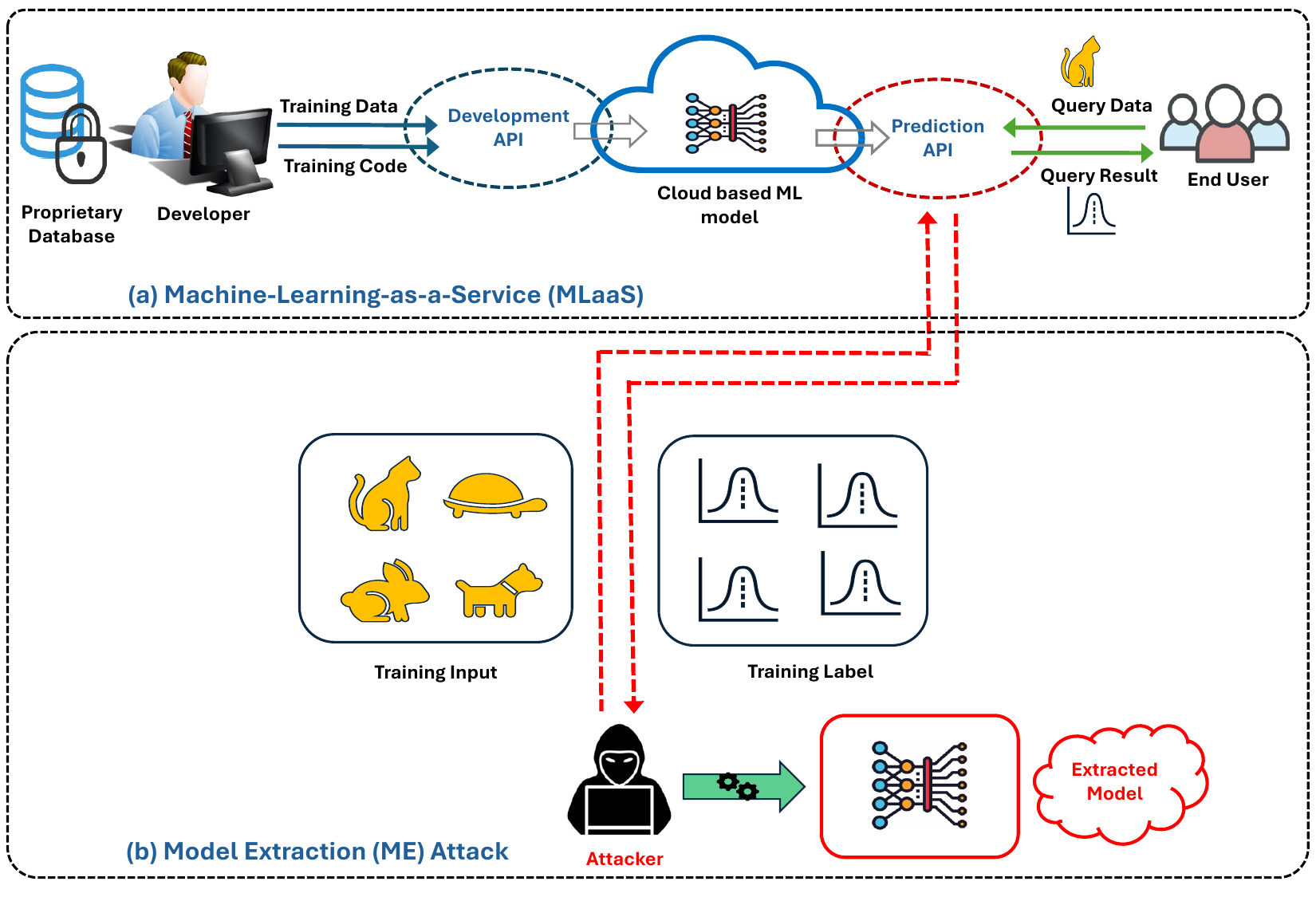}
    \caption{ME attack executed through predictive API in a Machine Learning as a Service (MLaaS) platform.}
    \label{fig:1}
    \vspace{-5pt}
\end{figure}

Although FL is intended to safeguard personal data, recent studies reveal that FL models are at risk of different attacks that can disclose sensitive information from training datasets \cite{hu2021source}, \cite{banerjee2024mia}. These vulnerabilities include model extraction (ME), reconstruction, membership inference, and model inversion (MI) attacks \cite{ma2020safeguarding}. An ME attack occurs when an adversary replicates the functionality of a victim model by querying an Application Programming Interface (API) without internal insight, leading to an extracted model that approximates the original model without direct access to its parameters or training data \cite{liang2024model}. 
In reality, an ME attacker amis for an approximate extraction that focuses on constructing an extracted or piracy model that closely resembles the victim model \cite{gong2021inversenet}, which can either achieve comparable performance to the victim model (measured by accuracy) or exhibit similar behavior (measured by fidelity). Figure \ref{fig:1} illustrates an overview of the ME attack process utilizing the predictive API on a MLaaS platform.

As FL models are deployed in MLaaS platforms, attackers can exploit predictive API queries to obtain insights about the victim model, effectively infringing on intellectual property (IP) and undermining the security measures intended to safeguard proprietary algorithms \cite{kesarwani2018model}. The collaborative exchange of model updates in FL can inadvertently expose sensitive information, making FL systems particularly susceptible to ME attacks, which threaten both data privacy and IP \cite{li2023model}. However, there has been only limited exploration of the impact of ME attacks in a scalable FL environment, particularly concerning accuracy and fidelity analysis, leaving this as an open area for further investigation. MLaaS platforms implement strong authentication mechanisms, such as API keys, OAuth, and multi-factor authentication, to restrict model access to authorized users \cite{vangala2022blockchain}. However, an authenticated adversary can still execute model extraction by issuing an excessive number of queries. In this paper, we investigate three key research questions. \textbf{First,} How vulnerable is an FL-based victim model to ME attacks, particularly with the size of the attack query set and the number of FL clients? \textbf{Second,} under an FL environment, how do various deep learning architectures (DL) affect the fidelity and accuracy of extracted models? \textbf{Finally,} how does the use of transfer learning (TL) with pre-trained models influence the effectiveness of ME attacks compared to models trained from scratch? 

The main contributions of this paper are as follows:

\begin{itemize}
    \item This paper evaluates the vulnerability of FL-based victim models to ME attacks, focusing on how query set size, FL clients, and deep learning architectures impact the fidelity and accuracy of the extracted models.

    \item  We investigate the use of pre-trained models as a starting point for extraction instead of training from scratch. This approach allows us to evaluate the impact of pre-trained models on attack accuracy and query efficiency, revealing potential advantages for attackers.
    
    \item We demonstrate that the TL-based ME attack approach enables the extraction model to surpass the best accuracy (training from scratch) with fewer query samples. This method allows the extraction model to closely replicate the victim model's performance, nearly matching its accuracy. 
\end{itemize}

The structure of this paper is as follows: \Cref{threatModel} outlines the threat model, detailing the attacker's objectives, knowledge, and capabilities. \Cref{overview} describes the framework for executing the ME attack within the FL environment. In \Cref{proposedAttack}, we introduce and explain the proposed algorithm for the ME attack in FL. \Cref{results} provides the experimental results along with an analysis and discussion of the findings. Lastly, we conclude our work and discuss a few future research thoughts in Section \Cref{conclusion}.

\section{Threat Model} \label{threatModel}

In this section, we outline the threat model, detailing the adversary's knowledge, goals, capabilities, and the scope of the proposed ME attack in FL \cite{tramer2016stealing}, \cite{jagielski2020high}. Machine learning models deployed in critical infrastructures are increasingly targeted by adversarial threats, including ME attack \cite{das2022securing}.

\textbf{Adversary’s objective:} The adversary $\mathcal{A}$ aims to create an extracted model $\mathcal{M}_e$ that closely replicates the functionality and/or performance of the MLaaS backend model, referred to as the victim model $\mathcal{M}_v$. The similarity between $\mathcal{M}_e$ and $\mathcal{M}_v$ is evaluated based on either accuracy or fidelity using a test dataset $\mathcal{D}^*$. The attacker does not modify the model’s parameters and requires no extra information \cite{liang2024model}.

\textbf{Adversary’s knowledge:} We consider a scenario where $\mathcal{A}$ possesses minimal information about the victim model $\mathcal{M}_v$, such as its architecture, hyperparameters, or the exact dataset used for training. However, the adversary does have access to an unlabeled reference dataset $\mathcal{D}$. While the training goals and model architecture are known to all FL participants, the adversary lacks any insight into the global training process (whether centralized or FL) and the distribution of the training data across clients.

\textbf{Adversary’s capability:} $\mathcal{A}$ can interact with $\mathcal{M}_v$ via the MLaaS API, which returns the prediction $\mathcal{M}_v(x)$ for any given input query $x$. These queries are not limited to real-world data and may also include synthetic or adversarial inputs. The attacker lacks the ability to modify the model's parameters and does not require additional information. Furthermore, as MLaaS typically operates on a pay-per-query basis, we assume the adversary is constrained by a limited query budget $n_{query}$ \cite{truong2021data}. $\mathcal{A}$ samples $n_{query}$ inputs query dataset $\mathcal{Q}$ to MLaaS prediction API and use all the query-response pairs $\{x,\mathcal{M}_v(x)\}$ to train the extracted model $\mathcal{M}_e$ by minimizing the cross-entropy loss \cite{liang2024model}.

\section{Framework for Proposed ME Attack in FL} \label{overview}

\subsection{Attack Overview}

In an ME attack scenario within an FL environment, $\mathcal{A}$ replicates a global victim model $\mathcal{M}_v$ by exploiting its MLaaS API. $\mathcal{A}$ submits API queries without direct access to the model's internal structure or training data. By collecting sufficient input-output pairs, $\mathcal{A}$ trains an extraction or surrogate model $\mathcal{M}_e$ that closely mimics $\mathcal{M}_v$. The attack success of $\mathcal{A}$ is determined by the $\mathcal{M}_e$’s fidelity, or how accurately it replicates the victim model’s behavior, and attack accuracy, reflecting how well its predictions align with the original model’s outputs. Even with restricted access, a well-executed query strategy can achieve high fidelity in FL and MLaaS setups.

\begin{figure*}[!htb]
    \centering
    \vspace{-15pt}
    \includegraphics[width=0.65\linewidth]{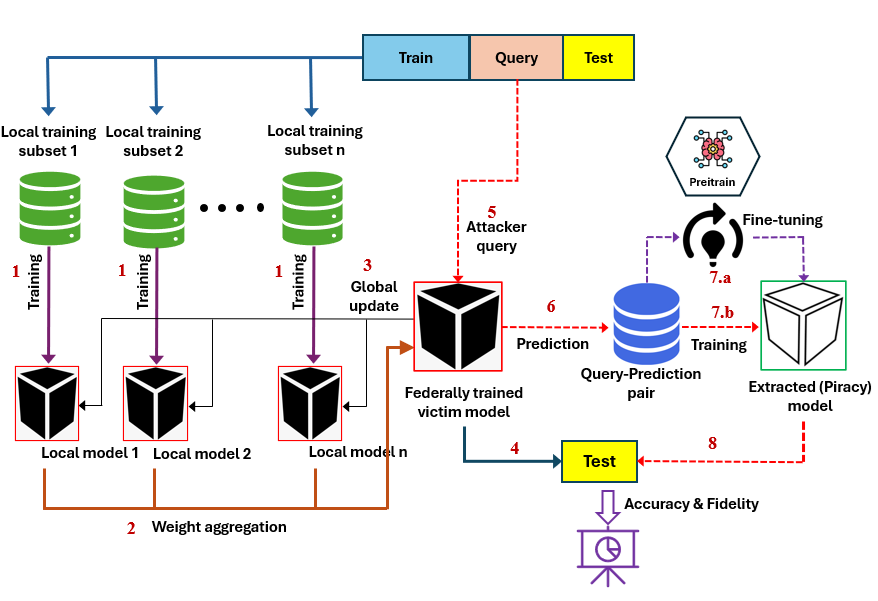}
    \vspace{-10pt}
    \caption{Framework of the proposed TL-based ME attack executed in an FL environment.}
    \label{fig:2}
    \vspace{-10pt}
\end{figure*}

\subsection{Victim Model Built in FL}

The victim model $\mathcal{M}_v$ in FL is the global model formed by aggregating updates from multiple participants. In MLaaS environments, this model is accessible through an API, allowing external users to query it for predictions without exposing its internal workings. The attacker leverages this API access to collect numerous input-output pairs. Despite lacking knowledge of the model’s architecture or training data, repeated queries enable the adversary to approximate the model's decision boundaries, ultimately building a highly accurate $\mathcal{M}_e$ that replicates the global model's predictive performance.

\subsection{Extracted (Surrogate) Model}

The extracted (surrogate) model $\mathcal{M}_e$ is the attacker’s replication of $\mathcal{M}_v$, trained on input-output pairs gathered from querying the global model. The fidelity of the surrogate indicates how closely it mirrors the original model, while attack accuracy measures its predictive performance against the victim model. Once trained, $\mathcal{A}$ can utilize $\mathcal{M}_e$ to bypass service restrictions or exploit the proprietary model’s functionality without authorization. $\mathcal{M}_e$ can also facilitate additional attacks, all while maintaining no direct access to the original training data or model parameters.

\section{Proposed ME attack in FL} \label{proposedAttack}
In this section, we explain the proposed TL-based ME attack in an FL environment. We first briefly describe the attack environment of datasets, extracted models, FL environments, and attack evaluation metrics. 

Figure \ref{fig:2} presents the proposed framework for executing a TL-based ME attack within a FL environment.

\textbf{Datasets} – In this experiment, we utilize three benchmark datasets: CIFAR-10, 
FashionMNIST, 
and MNIST
.Each dataset is partitioned into training and testing sets, where the test sets remain constant throughout the experiment to evaluate the performance of both the victim and extraction models. The training set is evenly split, with one half used to train the $\mathcal{M}_v$ and the other half designated for generating the query dataset used in the model extraction process.

\textbf{Extracted Models} – We consider two machine learning models, basic CNN and ResNet,
to evaluate the effectiveness of the ME attacks. In the default setting, the model parameters are randomly initialized, meaning the models are trained from scratch. To further enhance the performance of the ME attack, we also include pre-trained models, which are subsequently fine-tuned on the extraction dataset.

\textbf{FL Environments }– We design and implement an FL architecture to evaluate the ME attacks, incorporating two configurations: one with five clients and another with ten clients. In both setups, the FedAvg algorithm is employed to aggregate the model updates from the clients at the central server, creating a global model \cite{mcmahan2017communication}. For this experiment, we utilize the NVFlare (NVIDIA) library to develop and execute the FL architecture on GPU, ensuring efficient distributed training across the clients.

\textbf{Metrics} – To assess the effectiveness of ME attacks, we focus on three key metrics:

\textit{Accuracy} represents the proportion of inputs from the test set that are correctly classified by $\mathcal{M}_e$. 




\textit{Fidelity} quantifies the proportion of inputs from the test set that are classified identically by both $\mathcal{M}_v$ and $\mathcal{M}_e$. Formally,

\vspace{-0.2cm}
\begin{equation} \label{Eq.fidelity}
\textit{\text{Fidelity}}(\mathcal{M}_e) = \frac{\sum_{(x, y) \in \mathcal{D}^\star} \indicator\{\mathcal{M}_v(x) = \mathcal{M}_e(x)\}}{|\mathcal{D}^\star|}
\end{equation}

\textit{Kullback-Leibler (KL) divergence} quantifies the difference between two probability distributions. In ME attacks, it measures how the output probability distributions of $\mathcal{M}_v$ and $\mathcal{M}_e$ differ.

\subsection{The Overall Attack Paradigm} 


This subsection presents the fundamental steps and pseudo-code-based algorithm for the proposed TL-based ME attack implementation (Algorithm \ref{extraction_algo}).  

\begin{enumerate}
    \item Prepare the query dataset by generating a set of inputs. 

    \item Query $(\mathcal{M}_v)$ by sending each input $(q_i)$ to $(\mathcal{M}_v)$ and store the predictions $(p_i)$ for each query $(q_i)$ to create input-output pairs $(q_i, p_i)$. 

    \item Construct the extracted dataset by collecting all the input-output pairs $(q_i, p_i)$ gathered during the querying phase. This dataset $(D)$ serves as the training data for the extracted model, enabling it to learn and replicate the behavior of the victim model.

    \item Measure accuracy, fidelity, and KL divergence to validate the extracted model $(\mathcal{M}_e)$ by utilizing the test dataset $\mathcal{D}^*$ to evaluate the performance of both $(\mathcal{M}_v)$ and $(\mathcal{M}_e)$.
\end{enumerate}

We utilize the FL architecture to train the $\mathcal{M}_v$, while the extraction model is trained using conventional machine learning methods. First, the server initializing global model parameters and sending them to all clients. Each client updates the model locally with their data and sends the updated parameters back to the server, which aggregates these updates over multiple communication rounds to produce a trained global victim model.

\begin{algorithm}[!htb]
    \caption{TL-based ME Attack} \label{extraction_algo}
    \textbf{Input:} Query dataset $\mathcal{Q}$, Victim model $\mathcal{M}_v$, Test dataset $\mathcal{D}^*$, \mbox{\hspace{1cm}} Pre-train flag $pre\_train$\\
    \textbf{Output:} Extracted Model $\mathcal{M}_e$
    
    \begin{algorithmic}[1]
        \STATE Set $\mathcal{D}$ = [] \hspace{0.8cm} \textcolor{blue}{$\vartriangleright$ \textit{Extracted Dataset}}
        \STATE Initialize $\mathcal{M}_e \leftarrow$ Model Architecture

        \FOR{each $q_i$ in $\mathcal{Q}$}
            \STATE Send $q_i$ to the $\mathcal{M}_v$
            \STATE Record predictions $p_i \leftarrow \mathcal{M}_v(q_i)$
            \STATE Update $\mathcal{D} \leftarrow (q_i, p_i)$
        \ENDFOR
        
        \IF{$pre\_train$}
            \STATE Load pre-trained $\mathcal{M}_e$
            \STATE Fine-tune $\mathcal{M}_e$ on $\mathcal{D}$
        \ELSE
            \STATE Train $\mathcal{M}_e$ on $\mathcal{D}$
        \ENDIF
        
        \STATE Compute \textit{\text{Accuracy}}$(\mathcal{M}_e)$
        \STATE Compute \textit{\text{Fidelity}}$(\mathcal{M}_e)$, using Eq. (\ref{Eq.fidelity})
        \STATE Compute \textit{\text{KL Divergence}}$(\mathcal{M}_v \| \mathcal{M}_e)$

        \RETURN $\mathcal{M}_e$
    \end{algorithmic}
\end{algorithm}

The TL-based ME attack algorithm involves two primary stages: generating the query dataset and applying transfer learning presented in \Cref{extraction_algo}. In the first part (lines 1-7), the algorithm queries the \(\mathcal{M}_v\) using each sample from the query dataset \(\mathcal{Q}\), collecting the $\mathcal{M}_v$’s predictions to form an extracted dataset \(\mathcal{D}\). In the second part (lines 8-12), if pre-train is enabled, a pre-trained model is loaded and fine-tuned on the extracted dataset. Otherwise, the extraction model \(\mathcal{M}_e\) is trained from scratch. The algorithm then evaluates $\mathcal{M}_e$’s performance based on accuracy, fidelity, and KL divergence before returning \(\mathcal{M}_e\).

\section{Result Analysis and Discussion}\label{results}


This section presents the experimental setup, demonstrates the ME attack on both model types, and introduces an enhanced ME attack with pre-trained surrogate models and a TL approach.

\subsection{Experimental Setup}
The proposed ME attack is divided into two subsection: victim model training and extraction model training. For victim model training, we employ both centralized and federated training approaches. 

For the federated training approach, we outline an FL architecture designed to train ML models across various datasets. In FL architecture, the datasets are divided into $N$ distinct subsets and distributed into $N$ clients. Each client then trains a local instance of the model on its respective dataset. After local training, the clients securely transmit their model weights to a central server. The server aggregates these weights using a federated averaging method to create a global model, which is then redistributed to the clients \cite{mcmahan2017communication}. In subsequent rounds, the clients perform another epoch of local training on the updated global model and share the new weights with the server. This process is repeated for $T$ rounds to finalize the global model $\mathcal{M}_v$. 

\begin{table*}[ht]
	\centering
    \resizebox{0.9\linewidth}{!}{
		\begin{tabular}{|c|c|c|c|c|c|c|c|c|c|c|c|c|}
			\hline
			                                          &           \multirow{3}{*}{\textbf{Metrics}}            & \multirow{2}{*}{ \textbf{\shortstack{No. of\\ FL\\ Clients}}  } &                    \multicolumn{5}{c|}{\textbf{Basic CNN}}                     &                      \multicolumn{5}{c|}{\textbf{ResNet}}                      \\ \cline{4-13}
			                                          &                                               &                                                        & \multirow{2}{*}{\textbf{\shortstack{Victim \\ model}}} & \multicolumn{4}{c|}{\textbf{Extraction model}} & \multirow{2}{*}{\textbf{\shortstack{Victim \\ model}}} & \multicolumn{4}{c|}{\textbf{Extraction model}} \\ \cline{5-8}\cline{10-13}
			                                          &                                               &                                                        &                               & \textbf{ 5K}   &  \textbf{ 10K}   &  \textbf{ 20K}   &    \textbf{25k}    &                               &  \textbf{5K}   &  \textbf{10K}  &  \textbf{20K}  &      \textbf{25K}      \\ \cline{1-8}\cline{9-13}
			                                             & \multirow{3}{*}{\textbf{\shortstack{Accuracy\\(\%)} }} &Centralized       & 80.19 &59.99  &63.52  &70.72  &73.7       &76.22	&52.40	&58.73	&65.24	&68.36      \\ \cline{3-13}
			  \multirow{8}{*}{\rotatebox{90}{\textbf{CIFAR-10}}}  &   &5	&79.98	&56.49	&63.75	&70.67	&71.86	&81.05	&52.55	&58.65	&65.9	&68.07              \\ \cline{3-13}
			                                                      &   &10	&76.57	&53.13	&63.83	&70.05	&73.05	&82.39	&52.12	&59.19	&66.03	&68.94               \\ \cline{2-13}
                                                                                                    
			                                          &\multirow{3}{*}{\textbf{\shortstack{Fidelity\\(\%)}}}  &Centralized  &\multirow{3}{*}{ N/A }     &61.59  &62.44  &71.38   &73.52     &\multirow{3}{*}{ N/A }     &51.74	&57.9	&63.44	&66.87      \\ \cline{3-3}\cline{5-8}\cline{10-13}
			                                                         &      &5     &        &58.3	&65.02	&71.67	&72.57          &      &52.74	&58.71	&66.04	&68.15      \\ \cline{3-3}\cline{5-8}\cline{10-13}
			                                                         &      &10     &       &56.38	&65.23	&70.16	&73.4          &      &52.82	&58.82	&66.64	&68.38      \\ \cline{2-13}
                                                           
			                                          & \multirow{3}{*}{\textbf{\shortstack{KL\\Divergence}}}  &Centralized      &\multirow{3}{*}{ N/A }      &0.0322	&0.02744	&0.02265	&0.02069   &\multirow{3}{*}{ N/A }    &0.000486  &0.000416   &0.000348   &0.000318   \\ \cline{3-3}\cline{5-8}\cline{10-13}
			                                         &      &5     &        &0.0344	&0.02837	&0.022459	&0.01899          &      &0.000567	&0.000458	&0.000374	&0.0003436      \\ \cline{3-3}\cline{5-8}\cline{10-13}
			                                          &      &10     &       &0.03076	&0.02349	&0.01863	&0.0198          &      &0.000577	&0.000493	&0.000368	&0.000347      \\ \hline   \hline

                                                    & \multirow{3}{*}{\textbf{\shortstack{Accuracy\\(\%)} }} &Centralized       &98.94	&98.53	&99.06	&99.33	&99.39       &99.08	&98.36	&98.57	&99.17	&99.24      \\ \cline{3-13}
			 \multirow{8}{*}{\rotatebox{90}{\textbf{MNIST}}}   &   &5	&99.46	&98.62	&99.02	&99.12	&99.33	&99.46	&97.83	&98.76	&99.12	&99.01              \\      \cline{3-13}
			                                          &   &10	&99.39	&98.68	&99.14	&99.37	&99.32	&99.48	&97.48	&99.04	&99.02	&99.17               \\ \cline{2-13}
                                             
			                                          & \multirow{3}{*}{\textbf{\shortstack{Fidelity\\(\%)}}}  &Centralized  &\multirow{3}{*}{ N/A }     &98.48	 &98.82	 &98.85	 &98.87     &\multirow{3}{*}{ N/A }     &97.25	&97.87	&98.45	&98.59      \\ \cline{3-3}\cline{5-8}\cline{10-13}
			                                          &      &5     &        &98.58	&99.05	&99.2	&99.3          &     &97.51	&98.79	&98.41	&98.98      \\ \cline{3-3}\cline{5-8}\cline{10-13}
			                                           &      &10     &       &98.70	&99.04	&99.3	&99.3          &      &97.61	&98.83	&98.52	&98.79      \\ \cline{2-13}
                                              
			                                         & \multirow{3}{*}{\textbf{\shortstack{KL\\Divergence}}}  &Centralized      &\multirow{3}{*}{ N/A }      &0.0013	&0.001078	&0.00103	&0.0011697   &\multirow{3}{*}{ N/A }    &3.60E-05	&2.62E-05	&1.77E-05	&1.52E-05   \\ \cline{3-3}\cline{5-8}\cline{10-13}
			                                           &      &5     &       &0.00146	&0.000926	&0.00075	&0.000577          &     &3.04E-05	&1.54E-05	&2.35E-05	&1.18E-05      \\ \cline{3-3}\cline{3-3}\cline{5-8}\cline{10-13}
			                                           &      &10     &      &0.00129	&0.000919	&0.00056	&0.000572          &     &3.0379E-05	&2.38E-05	&1.72E-05	&1.26E-05      \\ \hline   \hline

                                                    & \multirow{3}{*}{\textbf{\shortstack{Accuracy\\(\%)} }}  &Centralized      &92.01	&88.16	&88.7	&90.74	&91.23	&89.65	&83.62	&86.94	&88.98	&89.49      \\  \cline{3-13}
		\multirow{8}{*}{\rotatebox{90}{\textbf{Fashion-MNIST}}}  &   &5	&92.21	&87.96	&89.57	&91.22	&91.39	&91.78	&85.30	&87.25	&89.43	&88.88              \\  \cline{3-13}
			                                           &   &10	&91.58	&87.64	&89.31	&91.04	&91.16	&91.85	&85.97	&87.03	&89.29	&89.7              \\ \cline{2-13}
                                              
			                                          & \multirow{3}{*}{\textbf{\shortstack{Fidelity\\(\%)}}}  &Centralized  &\multirow{3}{*}{ N/A }     &89.67	&90.83	&91.68	&92.05     &\multirow{3}{*}{ N/A }    &83.15	&86.85	&88.12	&89.01      \\ \cline{3-3}\cline{5-8}\cline{10-13}
			                                          &      &5     &        &89.53	&91.29	&92.49	&92.72          &      &86.83	&87.2	&89.71	&89.95      \\ \cline{3-3}\cline{5-8}\cline{10-13}
			                                                         &      &10     &      &89.88	&91.03	&91.5	&91.48          &      &85.99	&87.09	&89.66	&89.65      \\ \cline{2-13}
                                             
			                                          & \multirow{3}{*}{\textbf{\shortstack{KL\\Divergence}}}  &Centralized      &\multirow{3}{*}{ N/A }     &0.0135	&0.006298	&0.00677	&0.00628   &\multirow{3}{*}{ N/A }   &0.000125	&9.50E-05	&8.84E-05	&8.71E-05   \\ \cline{3-3}\cline{5-8}\cline{10-13}
			                                           &      &5     &        &0.00789	&0.006897	&0.005633	&0.00572          &     &0.000134	&1.20E-04	&1.05E-04	&9.13E-05      \\ \cline{3-3}\cline{5-8}\cline{10-13}
			                                          &      &10     &      &0.00693	&0.00653	&0.006176	&0.00578          &     &0.0001416	&1.14E-04	&9.32E-05	&9.23E-05      \\ \hline
			                                         
		\end{tabular}
	}
	\caption{ME atatck accuracy and fidelity across various FL clients on CIFAR-10, MNIST, and FashionMNIST datasets, evaluated on different DL model architectures.} \label{table1} 
    \vspace{-12pt}
\end{table*}

For the extraction (or surrogate) model, we utilize four distinct sample query datasets, consisting of 5k, 10k, 15k, and 20k queries. These query sets are used to generate query-prediction pairs, which are subsequently used to train $\mathcal{M}_e$. In this study, we mainly focus on three datasets- CIFAR-10, MNIST, and FashionMNIST, and two deep learning (DL) models, basic CNN and ResNet to evaluate the ME attack. Initially, $\mathcal{M}_e$ is trained from scratch without utilizing pre-trained model parameters. We then present a TL-based ME attack approach, where pre-trained model parameters are utilized for $\mathcal{M}_e$ to enhance the ME attack performance.

\subsection{ME Attack without Pre-trained Model}
In this section, we demonstrate the effectiveness of the ME attack on machine learning models trained using both centralized and federated approaches with 5 and 10 clients. \Cref{table1} summarize the accuracy, fidelity, and KL divergence of the ME attack for CIFAR-10, MNIST, and FashionMNIST. The accuracy of $\mathcal{M}_v$ serves as the baseline accuracy, which we aim to closely match. In this experiment, we apply both centralized and federate training approaches to the victim model, while for $\mathcal{M}_e$, we exclusively employ centralized training. 

For example, in the case of the CIFAR-10 dataset using a basic CNN model, the baseline accuracy of $\mathcal{M}_v$ is approximately 80.19\%. Our objective is to closely approach this baseline accuracy in the extraction model. We observe that the accuracy of $\mathcal{M}_e$ is directly correlated with the size of the query set. A query dataset consisting of 25k samples typically yields the highest $\mathcal{M}_e$ accuracy in both centralized and federated architectures. Similarly, for the ResNet model, accuracy is also strongly influenced by the size of the query set.

Similar trends are observed across other datasets, such as FashionMNIST and MNIST, where accuracy improves consistently with increasing query set size for both the CNN and ResNet models. Moreover, the $\mathcal{M}_v$'s performance also affects the accuracy of the extraction model. For instance, in the FashionMNIST dataset, the highest baseline accuracy is achieved using the basic CNN model with training distributed across five clients, which consequently results in the best extraction model accuracy for that setup.

When comparing the accuracy between the basic CNN and ResNet models, the basic CNN outperforms the ResNet model on both the CIFAR-10 and FashionMNIST datasets. However, when employing a pre-trained ResNet model, we achieve significantly better results (as discussed later).

Since the MNIST datasets are less complex than CIFAR-10, the extraction model can achieve high accuracy with fewer query samples. For example, using the FashionMNIST dataset with only a 5k query set, we achieve an extraction accuracy of approximately 88.16\% with the basic CNN model, compared to a baseline accuracy of around 92\%. However, under the same conditions with the ResNet model, the accuracy is significantly lower. Thus, we hypothesize that the extraction model's accuracy is strongly influenced by both the size of the query set and the performance of $\mathcal{M}_v$.

To further evaluate the ME attack's effectiveness, we assess fidelity and KL divergence to measure how well the extraction model approximates $\mathcal{M}_v$. Similar to accuracy, these metrics are influenced by the query dataset size, larger datasets generally yield higher extraction accuracy and, consequently, better fidelity and lower divergence. Since fidelity is inherently tied to extraction model accuracy, improvements in accuracy enhance fidelity, while reductions degrade both fidelity and KL divergence. These findings underscore the critical role of query set size and victim model performance in shaping ME attack success.



\renewcommand{\arraystretch}{1.5} 
\begin{table}[!htb]
\caption{ME attack performance using ResNet pre-trained model on CIFAR-10.} 
\label{table:cifar10_performance_accuracy_fidelity} 
\centering
\resizebox{\columnwidth}{!}{%
\begin{tabular}{| c | c | c | c | c | c | c | c | c | c |} 
        \hline
        \multirow{2}{*}{\rotatebox{90}{\textbf{Metric}}} & \multirow{2}{*}{\textbf{N}}  & \multicolumn{4}{c|}{\textbf{Training-from-scratch}} & \multicolumn{4}{c|}{\textbf{Using pre-trained model}} \\
        \cline{3-10}  
        &    & \textbf{5k}    & \textbf{10k}   & \textbf{20k}   & \textbf{25k}   & \textbf{5k}    & \textbf{10k}   & \textbf{20k}   & \textbf{25k} \\
        \hline \hline
        \multirow{3}{*}{\rotatebox{90}{\textbf{Accuracy}}} & 0   & 52.4  & 58.73  & 65.24  & 68.66  & 63.75  & 70.94  & 73.67  & 75.77  \\
                                           & 5   & 52.55  & 58.65  & 65.9  & 68.07  & 67.27  & 70.27  & 73.89  & 75.61  \\
                                           & 10  & 52.12  & 59.19  & 66.03  & 68.94  & 66.88  & 69.70  & 73.65  & 76.12  \\
        \hline
        \multirow{3}{*}{\rotatebox{90}{\textbf{Fidelity}}} & 0   & 51.74  & 57.9  & 63.44  & 66.87  & 62.52  & 68.57  & 70.69  & 70.88  \\
                                           & 5   & 52.74  & 58.71  & 66.04  & 68.15 & 68.3  & 70.3  & 73.67 & 74.69 \\
                                           & 10  & 52.82  & 58.82  & 66.64  & 68.38  & 68.98  & 69.94  & 73.45  & 74.16  \\
        \hline
\end{tabular}

}
\end{table}

\renewcommand{\arraystretch}{1.5} 
\begin{table}[!htb]
\vspace{-5pt}
\caption{ME attack performance using the ResNet pre-trained model on FashionMNIST.} 
\label{table:fashionmnist_performance_accuracy_fidelity} 
\centering
\resizebox{\columnwidth}{!}{%
\begin{tabular}{| c | c | c | c | c | c | c | c | c | c |} 
        \hline
        \multirow{2}{*}{\rotatebox{90}{\textbf{Metric}}} & \multirow{2}{*}{\textbf{N}}  & \multicolumn{4}{c|}{\textbf{Training-from-scratch}} & \multicolumn{4}{c|}{\textbf{Using pre-trained model}} \\
        \cline{3-10}  
        &    & \textbf{5k}    & \textbf{10k}   & \textbf{20k}   & \textbf{25k}   & \textbf{5k}    & \textbf{10k}   & \textbf{20k}   & \textbf{25k} \\
        \hline \hline
        \multirow{3}{*}{\rotatebox{90}{\textbf{Accuracy}}} & 0   & 83.62  & 86.94  & 88.98  & 89.49  & 87.95  & 88.86  & 89.73  & 90.44  \\
                                           & 5   & 85.3  & 87.25  & 89.43  & 88.88  & 87.72  & 88.41  & 90.41  & 90.17  \\
                                           & 10  & 85.97  & 87.03  & 89.29  & 89.7  & 88.33  & 89.24  & 90.27  & 90.8  \\
        \hline
        \multirow{3}{*}{\rotatebox{90}{\textbf{Fidelity}}} & 0   & 83.15  & 86.85  & 88.12  & 89.01  & 88.57  & 88.08  & 88.94  & 90.23  \\
                                           & 5   & 86.83  & 87.2  & 89.71  & 89.95 & 89.19  & 89.64  & 91.3 & 91.28 \\
                                           & 10  & 85.99  & 87.09  & 89.66  & 89.65  & 89.46  & 90.72  & 90.82  & 90.44  \\
        \hline
\end{tabular}
}
\vspace{-10pt}
\end{table}

\subsection{ME Attack with Pre-trained Model}
We leverage the TL approach to enhance the performance of the ME attack across multiple datasets \cite{ahamed2021attl}, \cite{ahamed2024targeted}. In this experiment, we employ a pre-trained ResNet model as the extraction model, which is subsequently fine-tuned on the extracted dataset. The ME attack performance for all query sets on the victim ResNet model, trained on the CIFAR-10 and FashionMNIST datasets, is presented in \Cref{table:cifar10_performance_accuracy_fidelity} and \Cref{table:fashionmnist_performance_accuracy_fidelity}, respectively. Across all query sets, the pre-trained model consistently surpasses the original extraction model in both accuracy and fidelity metrics.

For instance, in the CIFAR-10 dataset, the highest accuracy achieved by the basic CNN model with a 25k query set is around 73\%. However, when applying the TL approach on the same victim model, this accuracy is surpassed with a smaller query set (20k). The highest recorded extraction model accuracy for CIFAR-10 is around 76.12\%, closely matching the baseline accuracy of 76.52\%, effectively replicating the victim model's performance. A similar trend is observed in the FashionMNIST dataset, where TL leads in both accuracy and fidelity metrics across various query sets.

\begin{figure}[h!]
    \vspace{-5pt}
    \centering   
    \includegraphics[width=0.4\textwidth]{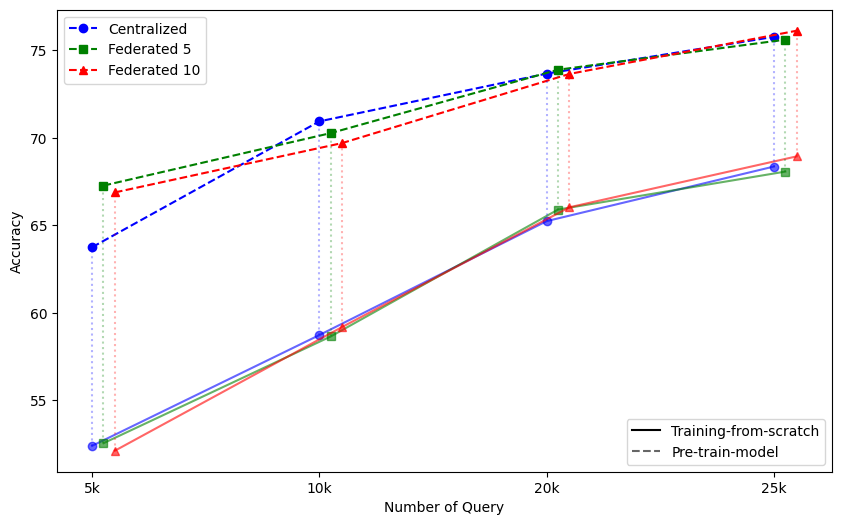}
    \caption{ME attack accuracy with ResNet pre-trained model on CIFAR-10.}
    \label{fig:ME attack accuracy with ResNet CIFAR10}

\end{figure}

\begin{figure}[h!]
    \vspace{-10pt}
    \centering   
    \includegraphics[width=0.4\textwidth]{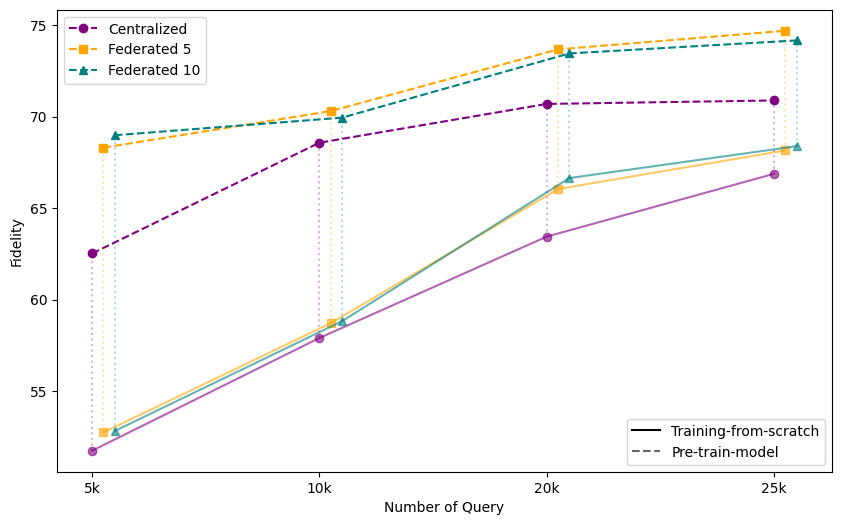}
    \caption{ME attack fidelity with ResNet pre-trained model on CIFAR-10.}
    \label{fig:ME attack fidelity with ResNet CIFAR10}
    \vspace{-10pt}
\end{figure}

\Cref{fig:ME attack accuracy with ResNet CIFAR10} and \Cref{fig:ME attack fidelity with ResNet CIFAR10} illustrate the notable improvements in both accuracy and fidelity for the CIFAR-10 dataset. These results clearly indicate that the performance of the ME attack is strongly tied to the extracted model's parameters. By incorporating pre-trained parameters, the extraction model achieves significantly better accuracy and fidelity, particularly with smaller query sets, thereby improving the overall effectiveness of the ME attack. For example, with a 10k query set on the CIFAR-10 dataset, the TL approach results in approximately 12.21\% higher accuracy compared to the original extraction model accuracy.

\section{Conclusion and future Scope}\label{conclusion}
In this study, we examine the vulnerability of FL-based models to ME attacks, showing that the accuracy and fidelity of extracted models are significantly affected by factors such as query set size, model architecture, and training datasets. Further, incorporating transfer learning (TL) into the extraction process notably enhances attack performance, especially with smaller query sets.  

In the future, we plan to focus on developing defenses against ME attacks, including techniques like noise injection in API responses and other stronger privacy-preserving methods in FL environments. Extending the research to other data types and exploring larger, more diverse federated networks will offer further insights into mitigating these risks.


\end{document}